\documentclass[12pt]{article}

\usepackage[letterpaper,margin=1in]{geometry}
\usepackage[T1]{fontenc}
\usepackage{lmodern}
\usepackage{microtype}
\usepackage[normalem]{ulem}
\usepackage{amsmath}
\usepackage{amssymb}
\usepackage{graphicx}
\usepackage{xcolor}
\usepackage{booktabs}
\usepackage{tabularx}
\usepackage{enumitem}
\usepackage{titlesec}
\usepackage{hyperref}
\usepackage{comment}

\hypersetup{
 colorlinks=true,
 linkcolor=black,
 urlcolor=black,
 citecolor=black
}

\definecolor{MDPrimary}{HTML}{1F4E5F} 
\definecolor{MDLight}{HTML}{F3F6F7}  

\setlist[itemize]{leftmargin=*,nosep}
\setlist[enumerate]{leftmargin=*,nosep}

\titleformat{\section}
 {\Large\bfseries\color{MDPrimary}}{\thesection}{0.6em}{}
\titleformat{\subsection}
 {\large\bfseries\color{MDPrimary}}{\thesubsection}{0.6em}{}
\titlespacing*{\section}{0pt}{1.2\baselineskip}{0.6\baselineskip}
\titlespacing*{\subsection}{0pt}{0.9\baselineskip}{0.4\baselineskip}

\newcommand{\MDTitle}[3]{%
 {\Huge\bfseries #1\par}%
 \vspace{0.3\baselineskip}%
 {\large #2\par}%
 {\normalsize \textcolor{MDPrimary}{#3}\par}%
 \vspace{0.8\baselineskip}\hrule\vspace{1.2\baselineskip}%
}

\begin{document}

\MDTitle
 {Topological cell-openness index for porous materials}
 {Michał Bogdan \quad\textbullet\quad Paweł Dłotko}
 {Institute of Mathematics of the Polish Academy of Sciences \quad\textbullet\quad mbogdan@impan.pl}

\begin{abstract}
We propose a method of estimating and parametrising the proportion of open and closed cells in a porous material based on measuring Betti numbers on the structures. We define a cell-openness index $\tau$ which can be used to complement the proportion of open-celled volume reported by gas pycnometry, which is the current gold standard for pore type characterization. We discuss in what types of structures mismatches between the two measures can occur and how such mismatches convey additional information about the structure. We demonstrate examples of significant correlations between $\tau$ and measurable physical quantities in both numerical and experimental structures. We also discuss how Betti curves can be used to estimate characteristic feature sizes in porous structures.
\end{abstract}

\noindent\textbf{\textcolor{MDPrimary}{Keywords:}}\ Porous materials \textbullet\ Open pores \textbullet\ Closed pores \textbullet\ Gas pycnometry \textbullet\ Topological data analysis \textbullet\ Persistent homology \textbullet\ Betti numbers \textbullet\ Microstructure characterization \textbullet\ 3D image analysis

\vspace{0.6\baselineskip}\hrule\vspace{0.8\baselineskip}

\section{Introduction}

The distinction between open- and closed-cell structures is widely used in the literature on cellular and porous materials. However, the two terms are not associated with a single, consistently applied operational criterion. In morphological descriptions, closed cells are commonly characterized by solid material occupying both the cell faces and edges, whereas in open cells only the edges remain solid \cite{Maire2003Cellular,Benedetti2021Architected,HossingerKalteis2021}. Other studies define the distinction primarily in terms of the pore phase: open cells are described as interconnected and accessible to fluid flow, while closed cells form mutually isolated pore regions \cite{KulshreshthaDhakad2020,Sriram2023Porous}. Yet another description distinguishes between continuous and discontinuous gaseous phases \cite{Duarte2020CellularMetals}. Some classifications additionally recognize an intermediate, semi-open class \cite{SunLi2018}, further indicating that cell openness is not intrinsically a binary property.

These descriptions are physically intuitive and largely consistent for idealised structures at the two ends of the open--closed spectrum. They are not, however, fully equivalent when applied to irregular experimental microstructures. Real materials may contain partially ruptured cell walls, a connected pore network coexisting with isolated pores, or several internally connected pore subnetworks that remain disconnected from one another. In such cases, classification as open-, semi-open-, or closed-cell may depend on the selected morphological or transport criterion, and the terminology does not by itself provide a reproducible quantitative measure of the degree of openness. Similarly, advanced two- and three-dimensional image analyses can quantify cell sizes, windows, strut lengths and other morphological characteristics \cite{Maire2003Cellular,Montminy2004Foams}, but these measurements do not directly produce a single scalar descriptor that places a structure along the continuum between the ideal open- and closed-cell cases. In particular, methods requiring each cell to be isolated as a separate connected voxel component are naturally applicable to closed-cell structures but do not extend directly to a connected open pore network.

This motivates the introduction of a quantitative, image-based convention for topological cell openness. In this work, we propose the cell-openness index $\tau$, which places porous structures on a continuous scale according to the topology of their pore phase. Within this convention, $\tau=0$ corresponds to an idealized structure composed of disconnected, simply connected pore components, characteristic of a closed-cell morphology. Values approaching $1$ correspond to pore phases containing relatively few connected components and many independent loops or tunnels, characteristic of highly interconnected, multiply connected open-cell networks. Intermediate values describe structures in which disconnected pore components and interconnected pore pathways coexist. The index is therefore intended not to replace every material-specific use of the terms open and closed, but to provide a reproducible topological measure that makes intermediate structures quantitatively comparable. In this respect,
it corresponds in range and basic interpretation to $\phi_0$- the fraction of open pore volume reported by gas pycnometry.

In this respect, $\tau$ corresponds in range and basic interpretation to the output of a traditional gas pycnometry experiment, which measures experimentally the fraction of the pores penetrable to an externally delivered gas and reports it as the "fraction of open pores" or "open porosity"\cite{Markl2018, Rouquerol1994, Webb2001}. Throughout this paper, we denote it as $\phi_0$ and use it as the benchmark against which $\tau$ is compared.

As any experimental method, gas pycnometry has its limitations.
Measuring $\phi_0$ relies on knowing the real density of the material \cite{Sereno2007}. Accuracy of the method may also be limited when otherwise open pores are not connected to the surface of the sample,
when sample size is not much larger than the pore size, and when some of the pores are blocked by contaminants such as condensating water microdroplets \cite{Nguyen2019}. This suggests the benefits of developing complementary
methods of parametrization of the degree of cell-openness in porous
materials, which are enabled by the rapid growth in accessibility of 3D
microimaging techniques \cite{Cnudde2013,Maire2014}.

Here, we believe that $\tau$ holds substantial promise, since, by construction, $\tau$ captures connectivity information that the existing standard for openness measurements (gas pycnometry) cannot resolve, most notably the existence of internally disconnected sub-networks within an otherwise ``open'' pore phase (see Sec. \ref{sec:methods_tau}).
 
Topologically describing porous and granular materials is not a new idea
in itself. Persistent homology has been applied to quantify pore similarity in nanoporous candidate materials for carbon capture or methane storage \cite{Lee2017}, uncover hidden hierarchical spatial structures in amorphous solids \cite{Saadatfar2017} and to estimate characteristic percolation-related properties within porous rocks \cite{Robins2016}. 

A related use of persistent homology in materials science concerns the
analysis of atomic configurations rather than pore networks. Hiraoka et al.~\cite{Hiraoka2016} used persistence
diagrams to characterize amorphous solids, where the absence of
long-range periodic order makes conventional structural descriptors,
such as pair correlation functions, insufficient for identifying
medium-range organization. Starting from atomic coordinates, they
considered unions of balls around atoms and tracked the resulting
topological features across scales: one-dimensional persistence
captures ring-like structures, while two-dimensional persistence
captures cavities. Importantly, the birth and death coordinates of
persistence points encode also metric
information about the size and geometry of these features. In silica
glass, this approach distinguished crystalline, liquid, and amorphous
configurations through qualitatively different persistence-diagram
patterns, including islands, diffuse clouds, and curve-like
distributions. This work demonstrated that persistent homology can
uncover hidden geometric organization in disordered materials and
relate it to physically meaningful structural signatures.

Compared to the aforementioned works, our main goal here is deliberately reductive: we ask
whether a single, interpretable number, directly comparable to
the one number returned by a pycnometer- can already capture the
open/closed pore distinction in a way that complements, rather than
replaces, the classical method of gas pycnometry. A secondary goal of this work is to
revisit the related question of recovering characteristic feature sizes
(pore radius, inter-pore distance, solid structure throat width) from the same TDA
signature, via simple geometric features of Betti curves, on our synthetically generated datasets (related earlier
work on recovering length scales from topological signatures is further
reviewed in Sec.~\ref{sec:results_lengthscales}.)

 This paper is organized as follows. Section~\ref{sec:datasets} describes the 2D and 3D microstructural image datasets used in this study. Section~\ref{sec:methods} introduces the numerical analogue of gas pycnometry, recalls the topological-data-analysis machinery (Betti numbers and Betti curves) that underpins our approach, defines the cell-openness index $\tau$, describes the heat-transfer and permeability simulation protocols, and details the statistical analysis. Section~\ref{sec:results} presents the results in three parts: (i)~the agreement and mismatch between $\tau$ and the fraction of open pores $\phi_0$ obtained using a numerical analogue of gas pycnometry, (ii)~length-scale estimates obtained from Betti curves in closed- and open-pore structures, and (iii)~the relationship of $\tau$ to permeability and effective thermal conductivity on selected datasets. Section~\ref{sec:discussion} discusses the implications, limitations and prospects of the proposed approach.

\section{Materials and microstructural image datasets}\label{sec:datasets}

All synthetic structures analysed in this study are voxelised binary microstructures stored in the \texttt{.npy} format: in the 2D case on a $400\times 400$ grid, in the 3D case on a $400\times 400\times 400$ grid (with the exception of the 3D dataset used for heat-transfer simulations, which uses a $200^3$ grid). In all cases, pore voxels are labelled $0$ and solid voxels are labelled $1$. The data is available together with the generating codes at 10.5281/zenodo.20751340.

\begin{figure}[ht]
 \centering
 \includegraphics[width=0.9\linewidth]{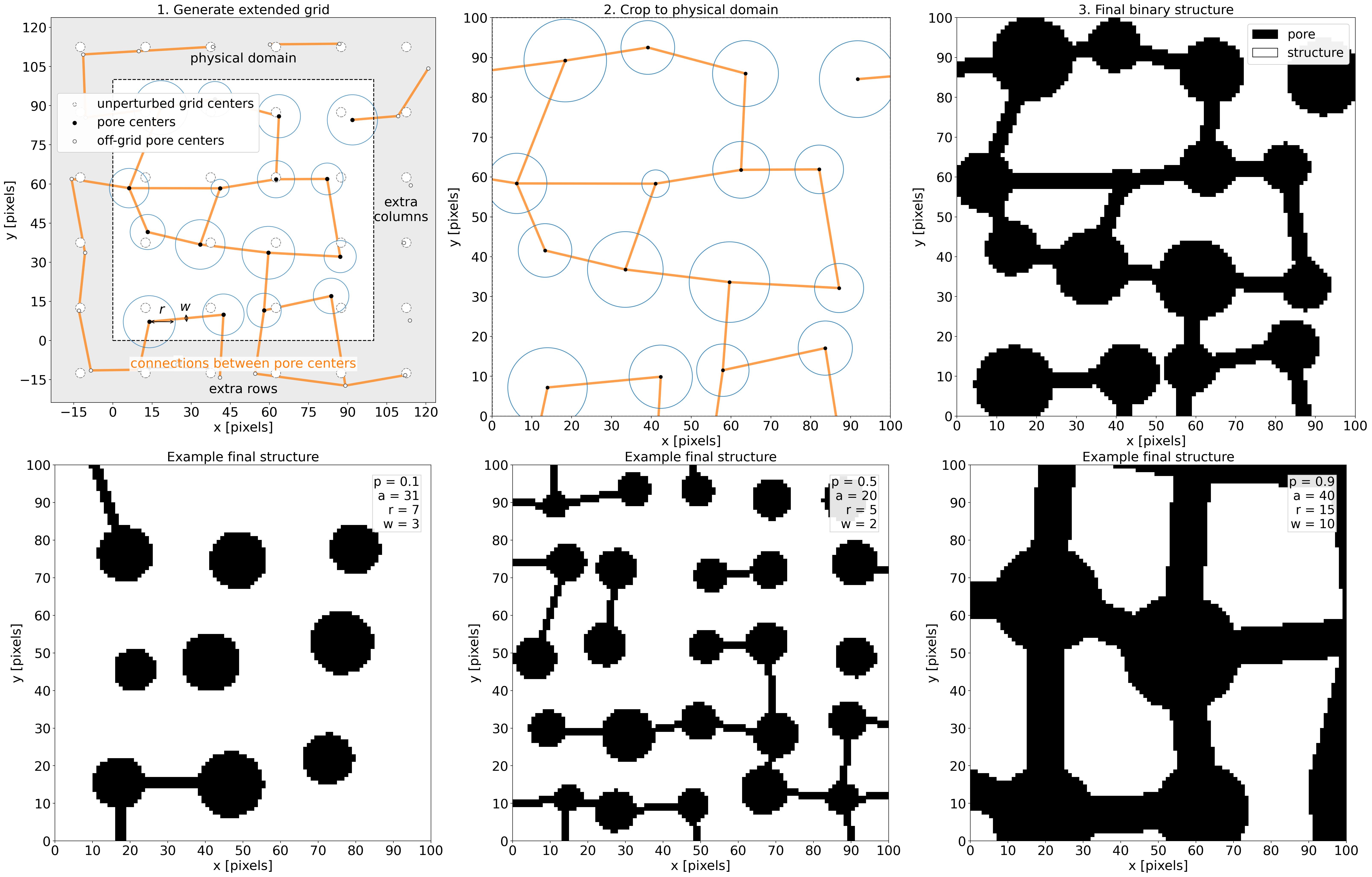}
 \caption{Schematic illustration of the synthetic porous-structure generation procedure. Pore centers are placed near a regular grid, pore radii and locations are randomised, and neighbouring pores are connected by channels according to the connection probability $p$.}
 \label{fig:porous_generation_scheme}
\end{figure}

\paragraph{Common construction of datasets generated for this work.} All families of synthetic structures generated for use in this work share the same basic features of the generative procedure. Each structure is otherwise solid, with circular (in 2D) or spherical (in 3D) pores of mean radius $r$ located close to nodes of a quadratic (in 2D) or cubic (in 3D) grid of mean spacing $a$. Each pore radius and each precise pore location is drawn independently from a normal distribution with standard deviations $\sigma_r$ and $\sigma_a$, respectively. The unit of $r$, $a$ and of all other distance-related parameters is dimensionless and equal to one pixel/voxel; the actual pores are obtained by performing a binary dilation operation around the pore center. Where channels connecting neighbouring pores are present, they have mean width $w$ with a standard deviation $\sigma_w$, and each pair of neighbouring pore centers is connected by such a channel independently with probability $p$ (and is left disconnected with probability $1-p$). The parameters $p$, $a$, $\sigma_a$, $r$, $\sigma_r$, $w$, $\sigma_w$ are separately drawn for every generated structure. To avoid artificially disconnecting the structure from the grid boundary, one additional row and column (or, in 3D, slab) of pore centers is temporarily created in the preparatory phase on each side outside of the final structure's boundaries and connected to the real pore centers according to the same rules that govern how other pore centers are connected to each other. The generation procedure is illustrated schematically in Fig.~\ref{fig:porous_generation_scheme}.

The families of structures differ only in (i)~the connection probability $p$, (ii)~whether the structure is 2D or 3D, and (iii)~the sampling ranges of the remaining parameters. These differences are summarised in the subsections below.

\paragraph{Externally obtained dataset for permeability correlation studies.} In addition to our own numerically generated datasets, we used a publicly available dataset of binarised X-ray tomography images of real sandstone rocks, for which flow permeabilities were experimentally and numerically measured \cite{Neumann2020}, to verify the relationship between $\tau$ and permeability.

\subsection{The dataset of partially open cells}\label{sec:partially-open-cells-dataset}

This dataset (used for studying correlations between $\phi_0$ and $\tau$) exists in both a 2D and a 3D variant. The 2D variant consists of 100 structures on a $400\times 400$ grid; the 3D variant consists of 100 structures on a $400\times 400\times 400$ grid. Both variants follow the common construction described above. The parameters $p$, $a$, $\sigma_a$, $r$, $\sigma_r$, $w$, $\sigma_w$ are drawn from uniform distributions with ranges summarised in Table~\ref{tab:dataset_parameter_ranges}; $p$ in particular spans the full range $[0,1]$ to ensure a broad coverage of partially-connected pore networks. This dataset is used in Sec.~\ref{sec:correlating_indices} to study correlations between $\phi_0$ and $\tau$.

\begin{table}[ht]
\centering
\caption{Sampling ranges for parameters in the partially-connected-holes dataset used for correlating $\phi_0$ and $\tau$.}
\label{tab:dataset_parameter_ranges}
\begin{tabular}{lll}
\hline
\textbf{Parameter} & \textbf{Distribution} & \textbf{Range / constraint} \\
\hline
$r$    & uniform       & $r \sim \mathcal{U}(6,20)$ \\
$w$    & conditional uniform & $w \sim \mathcal{U}(2,r)$ \\
$a$    & conditional uniform & $a \sim \mathcal{U}(3r,8r)$, accepted only if $a \leq 50$ \\
$p$    & uniform       & $p \sim \mathcal{U}(0,1)$ \\
$\sigma_r$ & conditional uniform & $\sigma_r \sim \mathcal{U}(0,r)$ \\
$\sigma_w$ & conditional uniform & $\sigma_w \sim \mathcal{U}(0,w/2)$ \\
$\sigma_a$ & conditional uniform & $\sigma_a \sim \mathcal{U}(0,a/2)$ \\
\hline
\end{tabular}
\end{table}

\subsection{The closed-cells dataset}\label{sec:the-closed-holes-dataset}

This 2D dataset of 100 structures on a $400\times 400$ grid is used to verify what information Betti curves convey about characteristic structure sizes in structures with purely closed pores. Compared with the common construction, the only essential difference is that no channels connecting the pores are added (equivalently, $p=0$, and the parameters $w$ and $\sigma_w$ are not used). The remaining parameters are drawn from the ranges summarised in Table \ref{tab:dataset_parameter_ranges}, with the exception that the condition that $a<50$ is lifted and the upper limit on $r$ is lowered to 10.

\subsection{The open-cells datasets}\label{sec:the-open-network-dataset}

Two 2D datasets of $100$ structures on a $400\times 400$ grid were used to verify what information Betti curves convey about characteristic structure sizes in structures with purely open pore networks. The first of them has only a single difference with the common construction, in that the connection probability is fixed at
\[
p=1,
\]
i.e. every pair of neighbouring pore centers is connected by a channel. The remaining parameters for that dataset are drawn from the ranges summarised in Table \ref{tab:dataset_parameter_ranges}, with the exceptions that the upper limit on $w$ is now placed at $2r$ and the condition that $a < 50$ is lifted.

However, a second open-cells dataset was also constructed with a lower level of noise in the properties of structures that are generated there. In that dataset, $\sigma_r$, $\sigma_w$ and $\sigma_a$ for each structure are drawn from $\mathcal{U}(0,r/10)$, $\mathcal{U}(0,w/10)$ and $\mathcal{U}(0,a/10)$ respectively, instead of the wider ranges described in Table \ref{tab:dataset_parameter_ranges}.

\subsection{The mixed-cells 3D datasets (used for correlating $\tau$ with heat-transfer properties)}\label{sec:dataset_for_heat_transfer}

These 3D mixed-cells datasets are used for analysing relationships between $\tau$ and the heat-transfer tensor. They in principle follow the common construction, but live on a $200^3$ voxel grid (rather than $400^3$) for computational tractability of the heat-transfer solves and the ranges of specific parameters are very different from the ones specified in Table \ref{tab:dataset_parameter_ranges}, which also leads to qualitative differences, such as different pore centers merging into larger pores. Five independent ensembles of 50 samples each ($250$ samples in total) were generated with different parameter ranges in order to span a broad range of mean filled volume fraction (from $\sim 0.31$ to $\sim 0.94$). The sampling ranges and the resulting filled-volume-fraction statistics for each ensemble are summarised in Table~\ref{tab:mixed_cells_heat_transfer_datasets}.

\begin{table}[ht]
\centering
\caption{Sampling ranges and filled-volume-fraction summaries for the 3D mixed-cells datasets used in the heat-transfer analysis. $p$ is uniformly sampled between $0$ and $1$ for all datasets.}
\label{tab:mixed_cells_heat_transfer_datasets}
\resizebox{\linewidth}{!}{%
\begin{tabular}{lrllllrr}
\hline
\textbf{Subset no.} & \textbf{$n$} & \textbf{$r$ range} & \textbf{$a$ sampling} & \textbf{$w$ sampling} & \textbf{Mean filled VF} & \textbf{Min} & \textbf{Max} \\
\hline

1 & 50 & $[10,30]$ uniform & $a \sim \mathcal{U}(2.5r,4r)$, $a \leq 120$       & $w \sim \mathcal{U}(a/2,a)$ & 0.500374 & 0.044949 & 0.881445 \\
2 & 50 & $[6,20]$ uniform & $a \sim \mathcal{U}(2r,4r)$, $a \leq 80$        & $w \sim \mathcal{U}(2,a)$  & 0.314494 & 0.223779 & 0.844121 \\
3 & 50 & $[6,20]$ uniform & $a \sim \mathcal{U}(3r,8r)$, resample until $a \leq 50$ & $w \sim \mathcal{U}(2,r)$  & 0.857354 & 0.619307 & 0.980551 \\
4  & 50 & $[6,20]$ uniform & $a \sim \mathcal{U}(5r,10r)$, $a \leq 80$        & $w \sim \mathcal{U}(2,a/3)$ & 0.929096 & 0.790034 & 0.994846 \\
5  & 50 & $[6,14]$ uniform & $a \sim \mathcal{U}(5r,10r)$, $a \leq 80$        & $w \sim \mathcal{U}(2,a/3)$ & 0.944074 & 0.848254 & 0.992209 \\
\hline
\end{tabular}%
}
\end{table}

\section{Methods}\label{sec:methods}

\subsection{Numerical gas pycnometry/ connected porosity}\label{sec:methods_pycnometry}

 The numerical analogue of gas pycnometry that we use throughout this work, denoted $\phi_0$, is defined as the volume fraction of pore voxels that belong to a connected component of the pore phase touching at least one face of the cubical (or, in 2D, square) image domain. Operationally, $\phi_0$ is computed by:
\begin{enumerate}
 \item Labelling all connected components of the pore phase, using 26-connectivity in 3D (8-connectivity in 2D);
 \item Identifying which components contain at least one voxel/pixel adjacent to the domain boundary;
 \item Summing the voxel/pixel counts of those components and dividing by the total pore-voxel/pore-pixel count.
\end{enumerate}
This quantity is the digital counterpart of the open-porosity fraction reported by gas pycnometry experiments.

\subsection{Betti numbers and Betti curves}\label{sec:methods_betti}

 For a binary $d$-dimensional image $X \subseteq \mathbb{Z}^d$, the Betti numbers $\beta_0(X), \beta_1(X), \beta_2(X)$ count, respectively, the number of connected components, the number of independent 1-dimensional loops, and (in 3D) the number of enclosed cavities. Throughout this work, we compute Betti numbers on \emph{sub-level sets} of the signed distance transform (SDT) of the binary structure: pore voxels are assigned negative SDT values and solid voxels positive SDT values, and the filtration parameter $t$ ranges over integer pixel distances. The \emph{Betti curve} $\beta_i(t)$ is then the function $t \mapsto \beta_i(\{x : \mathrm{SDT}(x) \leq t\})$; the SDT filtration and corresponding persistent-homology output are illustrated in Fig.~\ref{fig:sdt_and_ph_demonstration}.

Because distances between pixels are discrete, Betti numbers are evaluated only at integer values of $t$; gradients $\nabla \beta_i(t)$ are computed using finite differences on this integer grid. In the analyses on the open-pore datasets, prior to gradient computation we restrict to persistent topological features with persistence $\geq 1.5$, as obtained from the persistence diagram associated with the same filtration; this filters out small-scale noise arising from voxelisation artefacts.

\subsection{Definition of the cell-openness index $\tau$}\label{sec:methods_tau}
    
The cell-openness index $\tau$ proposed in this work is a number in $[0,1]$ that summarises the openness of the pore network through its homological signature and is intended to complement $\phi_0$, reported by gas pycnometry. The index is defined using the Betti numbers of the \emph{pore phase}. More precisely, let $P$ denote the pore space in the original binary image, equivalently the sublevel set of the signed distance transform evaluated at filtration level $t=-1$.

The zeroth Betti number $\beta_0(P)$ counts the connected components of the pore phase. These include both components connected to the boundary of the image and components completely enclosed by the solid material. In structures containing one principal externally connected pore network, additional connected components correspond to isolated closed cells. Consequently, a large value of $\beta_0(P)$ relative to the higher-dimensional Betti numbers is characteristic of a predominantly closed-cell structure.

The first Betti number $\beta_1(P)$ counts independent loops in the pore network. Such loops arise when pores and channels are connected through multiple alternative pathways and are therefore characteristic of open, multiply connected pore structures. A larger value of $\beta_1(P)$ indicates a greater degree of topological interconnection within the pore phase.

In three dimensions, the second Betti number $\beta_2(P)$ counts independent enclosed cavities in the pore phase. Geometrically, these correspond to isolated components of solid material that are completely surrounded by pore space, such as solid grains ``floating'' inside the pore network. This situation is expected to be uncommon when the solid phase forms a single connected skeleton, but $\beta_2(P)$ is retained in the definition to accommodate structures in which such isolated solid components occur. Note that $\beta_2(P) \equiv 0$ in 2D.

The cell-openness index is defined both for 2D and 3D images as

\[
\tau = 
\frac{\beta_1(P)+\beta_2(P)}
{\beta_0(P)+\beta_1(P)+\beta_2(P)}.
\] 

\begin{figure}[ht]
 \centering
 \includegraphics[width=0.9\linewidth]{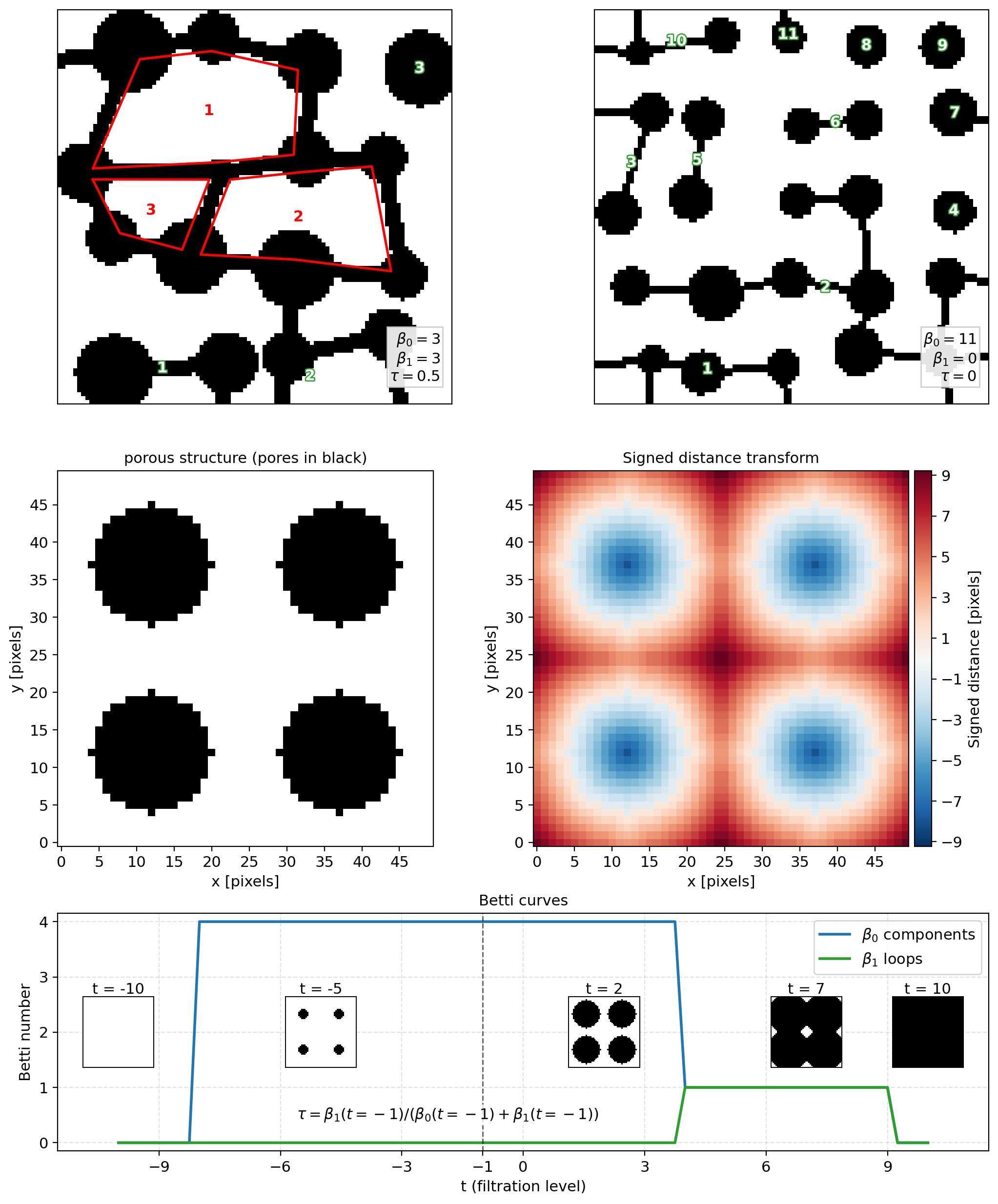}
 \caption{Demonstration of the workflow based on the signed distance transform and persistent homology used to compute the cell-openness index $\tau$.}
 \label{fig:sdt_and_ph_demonstration}
\end{figure}

The interpretation is particularly transparent when $\beta_2(P)=0$. A predominantly closed-cell structure contains many disconnected, approximately simply connected pore components. In this case, $\beta_0(P)$ is large relative to $\beta_1(P)$, and $\tau$ is close to $0$. Conversely, an open-cell structure typically contains one or a small number of connected pore components together with many independent loops or tunnels. In this case, $\beta_1(P)$ is large relative to $\beta_0(P)$, and $\tau$ approaches $1$. Thus, $\tau$ measures the relative predominance of interconnected, loop-forming pore architecture over disconnected pore components. The upper row of Fig. \ref{fig:sdt_and_ph_demonstration} demonstrates the values of Betti numbers and $\tau$ on examples of porous structures.

For pore networks containing relatively few connected components and many independent loops, $\tau$ approaches 1 while it is close to $0$ in a porous network in which all pores are disconnected from each other. Operationally, $\tau$ is obtained using the GUDHI library for Python~\cite{gudhi:urm,gudhi:CubicalComplex,gudhi:PersistentCohomology} by computing the values of Betti curves at filtration value of $-1$ (this is the value of SDT in porous voxels neighbouring voxels of the structure, while $SDT=0$ in voxels of the structure neighbouring the pores; therefore choosing a threshold of $-1$ as opposed to $0$ guarantees that the correct voxels get attributed to pores).
The SDT filtration also provides a natural framework for assessing the sensitivity of $\tau$ to image resolution and voxelisation. While $\tau(-1)$ describes the topology of the original segmented pore space, values at neighbouring filtration levels correspond to small erosions or dilations of that space. Consequently, the local variation of $\tau(t)$ around $t=-1$ can be used to quantify the robustness of the index to perturbations on the scale of one or a few voxels, which in real images frequently stems, for example, from the non-obvious selection of segmentation thresholds.

A schematic demonstration of the signed-distance-transform filtration and the corresponding persistent-homology output is shown in the middle and lowest rows of Fig.~\ref{fig:sdt_and_ph_demonstration}.

We have also developed a simpler implementation that computes $\tau$ directly on the original segmented pore space, without constructing the SDT filtration or computing persistent homology. For a three-dimensional binary image, the method first labels the connected components of the pore phase using 26-connectivity, which directly gives $\beta_0(P)$. It then labels the connected components of the complementary solid phase using the dual 6-connectivity; after excluding the component connected to the exterior of the domain, the remaining components correspond to enclosed cavities and therefore determine $\beta_2(P)$. Finally, using the Euler characteristic
\[
\chi(P)=\beta_0(P)-\beta_1(P)+\beta_2(P),
\]
the first Betti number is recovered as
\[
\beta_1(P)=\beta_0(P)+\beta_2(P)-\chi(P).
\]
These three Betti numbers are sufficient to evaluate the full cell-openness index $\tau$. This direct implementation, available together with the other codes at 10.5281/zenodo.20751340, was used on selected structures to cross-validate our principal method and is substantially faster. Unlike the SDT- and persistent-homology-based implementation, however, it does not allow us to assess the robustness of the index under small erosions, dilations, or segmentation perturbations, nor does it provide the filtration-dependent information used to estimate characteristic length scales of porous structures (see Sec.~\ref{sec:results_lengthscales}).

\subsection{Correlating $\tau$ with heat-transfer and permeability properties}\label{sec:methods_simulations}

\paragraph{Effective thermal conductivity (3D).}
For each of the $250$ structures in the datasets described in Sec.~\ref{sec:dataset_for_heat_transfer}, we computed the trace of the apparent/effective thermal-conductivity tensor $tr(K)$ by solving the steady diffusion equation
\[
\nabla \!\cdot (k \nabla T) = 0
\]
on the voxel grid with piecewise-isotropic conductivity $k=20$ in pores and $k=0.2$ in the solid. $tr(K)$ for each structure is obtained by solving for the steady-state temperature profile and heat flux after imposing Dirichlet temperatures on two opposing faces of the structure's grid ($T=1$ on one and $T=0$ on the opposite face) and zero normal flux on the remaining four faces. This procedure is separately done 3 times while swapping the faces on which the Dirichlet boundary conditions are imposed ($xy$, $yz$ and $xz$). Each simulation is initiated with a linear temperature gradient between the faces with Dirichlet temperature conditions. The Jacobi iterative method with a fixed iteration budget of 1200 iterations is used to equilibrate the temperature profile \cite{leveque2007} (the iteration budget allowed for a convergence below $10^{-4}$ for all 250 structures). We implemented the solution directly in python using the package numpy. The codes are available at 10.5281/zenodo.20751340.

\section{Results}\label{sec:results}

\subsection{Agreement and mismatch of $\tau$ with output of numerical pycnometry}\label{sec:correlating_indices}

\begin{figure}[ht]
 \centering
 \includegraphics[width=0.9\linewidth]{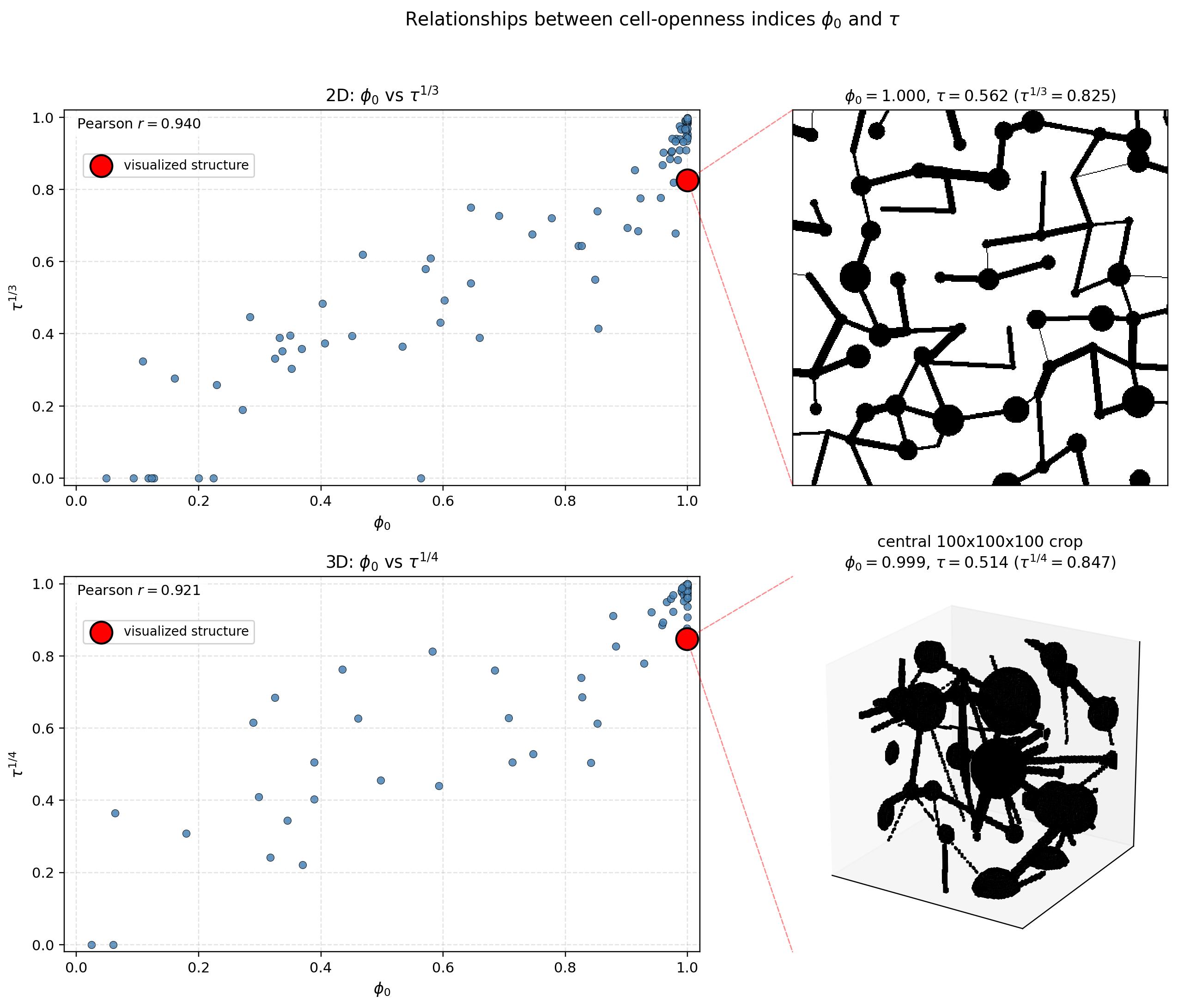}
 \caption{Correlations between cell-openness indices in 2D and 3D. Please note that pores are denoted as black in the insets, demonstrating example structures with a mismatch between $\tau$ and $\phi_0$.}
 \label{fig1:correlating_indices}
\end{figure}

We observe that $\tau$ and $\phi_0$ are highly correlated, corresponding to their roles of being complementary measures of the openness of pores in the material. The relationship appears to be a power law of the form $\phi_0 \propto \tau^{\frac{1}{k}}$, where $k$ is an integer. On the 2D variant of the dataset of partially open cells (see. Sec. \ref{sec:partially-open-cells-dataset}), the Pearson correlation between $\phi_0$ and $\tau^{\frac{1}{k}}$ is highest for $k=2$ and $k=3$ (respectively $0.933$ and $0.94$) and is lower with other root powers, while for the 3D version of that dataset it is maximised for $k=4$ (at $0.921$), with results for $k=3$ and $k=5$ trailing closely behind (respectively $0.921$ and $0.917$). The observations were carried out on datasets of 100 structures each for both the 2D and 3D case (see Sec.~\ref{sec:partially-open-cells-dataset} and Table~\ref{tab:dataset_parameter_ranges} for details on the dataset generation). The corresponding scatterplots are shown in Fig.~\ref{fig1:correlating_indices} (upper panel for 2D, lower panel for 3D). We observe in both cases substantial populations of structures in the upper-right corners of the scatterplots, for which $\tau$ is much more discriminatory than $\phi_0$ (for example, in 2D 16 out of 100 structures satisfy $\phi_0 > 0.98$ and $\tau^{1/2} < 0.95$). This suggests that within datasets of largely open-celled porous materials, indistinguishable using traditional gas pycnometry and $\phi_0$, $\tau$ presents an opportunity to discriminate between truly well-connected networks and those with weaknesses in their connectivity structure. An example of such a weakness captured by the mismatch between the two parameters is a pore network consisting of several components which, though almost all individually connected to the edge of the domain, are mutually disconnected- see the right column in Fig.~\ref{fig1:correlating_indices}.

\subsection{Betti-curve based length-scale estimates}\label{sec:results_lengthscales}

The idea that topological summaries of a porous structure can be used to recover its characteristic length scales has a long history. The classical approach is to track Minkowski functionals- volume, surface area, integral of mean curvature, and the Euler-Poincaré characteristic- under successive erosions or dilations of the binary structure, an idea introduced into statistical physics by Mecke \cite{Mecke2000} and now established as a standard family of descriptors for porous media \cite{Arns2010,Armstrong2019}, including soil structures \cite{Vogel2010}. Scholz et. al \cite{Scholz2012} demonstrated that the Euler characteristic alone is sufficient to predict the permeability within a certain class of numerically generated model porous materials. Robins et. al \cite{Robins2016} investigated how the persistence of topological signatures relates to percolation radii in porous materials, while Lautensack~et~al. \cite{Lautensack2008} used the evolution of the Euler number under iterative ball-erosion of a ceramic-foam micro-CT image to extract estimates for minimal and maximal strut thicknesses in open foams.

The approach we take here can be seen as a further exploration of this programme. Instead of tracking the Euler characteristic, which is an alternating sum of Betti numbers, we separately track the individual Betti curves $\beta_0(t)$ and $\beta_1(t)$ over the signed-distance filtration, and read off characteristic length scales from where each of them rises and falls. As we show below, this separation allows several distinct length scales to be recovered in closed-cell foams from the same filtration: not only the strut thickness, but also the pore radius and the inter-pore distance.

The idea behind our method is conceptually simple: as the filtration parameter $t$ moves from negative to positive values, the binary structure is progressively dilated, and topological features (connected components, loops) appear and disappear at filtration levels that correspond to natural length scales of the geometry. Tracking where $\beta_0$ and $\beta_1$ rise and fall therefore provides a dictionary of candidate structure property predictors.

\begin{figure}[ht]
 \centering
 \includegraphics[width=0.8\textwidth]{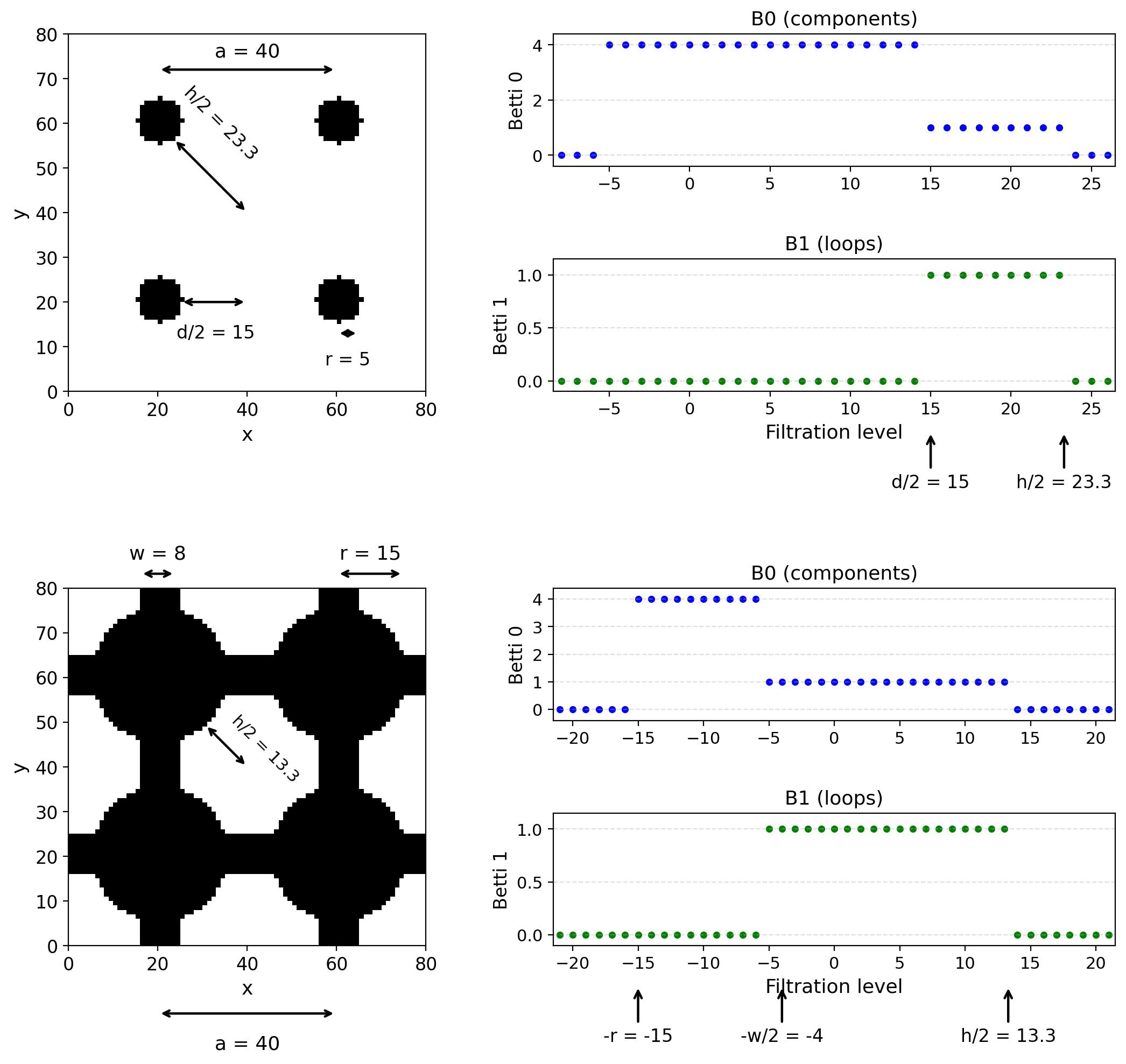}
 \caption{Illustrative Betti-curve features for an idealised closed-cell system (top) and an idealised open-cell system (bottom). Arrows annotate the filtration values at which $\beta_0$ and $\beta_1$ jump, together with the characteristic length scales ($r$, $d/2$, $w/2$, $h/2$) that those jumps recover.}
 \label{fig:shape_vs_Betti_pedagogical}
\end{figure}

We illustrate this dictionary in Fig.~\ref{fig:shape_vs_Betti_pedagogical} on two structures (one with closed-cell pores and one with open-cell pores): the closed-cell structure has isolated pores of mean radius $r$ on a square lattice of mean spacing $a$ (top row), and the open-cell structure has circular pores with the same symbols used for mean radii and spacings, but connected by throats of width $w$ (bottom row). In both cases, we propose to recover the following geometric quantities based on Betti curves:

\begin{itemize}
 \item \textbf{Mean pore radius $r$.} The pores in both the open- and the closed-cell system are counted by $\beta_0$ as separate connected components from the point when the filtration level $t = -r$; therefore, we propose that the point where $\nabla\beta_0$ is maximum estimates $r$ in both open-celled and closed-celled networks.
 \item \textbf{Inter-pore halfwidth $d/2$ (closed cells) / throat halfwidth $w/2$ (open cells).} As $t$ increases through positive values, neighbouring pores in a closed-cell structure ``merge'' through the gaps between them at $t = \frac{d}{2} = \frac{a}{2}- r$; therefore, $\beta_0$ drops and $\beta_1$ rises at filtration level $t=\frac{d}{2}$. In open-cell systems the merging of the pores with growing $t$ and the subsequent drop in $\beta_0$ and growth in $\beta_1$ occurs not at $\frac{d}{2}$ but much sooner, at negative filtration values- specifically, at negative half the throat width $-\frac{w}{2}$. Therefore, we propose three candidate predictors for inferring inter-pore halfwidth $\frac{d}{2}$ (in closed cells) and throat halfwidth $\frac{w}{2}$ (in open cells) from the Betti curves: (1) $\arg\min_t\nabla\beta_0(t)$, (2) $\arg\max_t\nabla\beta_1(t)$, and (3) the midpoint between them.
 \item \textbf{Half-thickness of the solid core $h/2$.} As $t$ grows further, the solid framework itself is eroded by the filtration, and last loops summed by $\beta_1$ close when the solid struts disappear at filtration level $t = h/2 = \sqrt{2}\,a/2 - r$. We therefore suggest that the point at which $\nabla\beta_1$ is maximally negative estimates $h/2$, both for closed- and open-cell structures. However, it is to be noted that for open-cell structures, depending on the relative magnitudes of $a$, $r$ and $w$, the equation for $h/2$ is in principle of the more general form of $t = h/2 = min(\sqrt{2}\,a/2 - r,a/2-w/2)$.
\end{itemize}

The two subsections below test the proposed predictors quantitatively against ground-truth simulation parameters.

\subsubsection{Closed-pore 2D structures}\label{sec:results_lengthscales_closed}

We test the dictionary of Sec.~\ref{sec:results_lengthscales} on the dataset of 100 randomised $400\times 400$-voxel structures with realistic noise and with closed pores described in Sec.~\ref{sec:the-closed-holes-dataset}. For each structure, the ground-truth length scales are $r$, $\frac{d}{2}=\frac{a}{2}-r$, and $\frac{h}{2}=\sqrt{2}\,\frac{a}{2}-r$. From its Betti curves we extract the four candidate predictors of Sec.~\ref{sec:results_lengthscales}. As noted in Sec.~\ref{sec:methods_betti}, Betti numbers are computed only at integer values of the filtration, and gradients are therefore evaluated on the same integer grid.

We perform a linear fit of each predictor against its corresponding target, with the intercept fixed at zero. The slope should be close to $1$ if the predictor matches the target quantity. We find that $\frac{d}{2}$ is reasonably well predicted by all three of its candidate predictors, but best predicted by the midpoint $\frac{1}{2}\!\left(\arg\min_t \nabla \beta_0(t)+\arg\max_t \nabla \beta_1(t)\right)$, while $r$ and $\frac{h}{2}$ are predicted with similar accuracy by their respective predictors. We note that for the relationship between $r$ and $-\arg\max_t \nabla \beta_0(t)$ the quality of the fit, as demonstrated by a relatively lower $R^2$, is lower than for other predictors, which apparently stems from the discretisation of the $x$-variable. Results are summarised in Table~\ref{tab:exp3_regressions}.

\begin{table}[ht]
\centering
\caption{Linear regressions (intercept fixed at $0$) between Betti-curve-derived predictors and geometric targets on the closed-cells dataset.}
\label{tab:exp3_regressions}
\begin{tabular}{llllr}
\hline
\textbf{target} & \textbf{predictor} & \textbf{slope} & \textbf{$R^2$} & \textbf{$n_{\mathrm{samples}}$} \\
\hline
$r$    & $-\arg\max_t \nabla \beta_0(t)$ & $0.985$ & $0.672$ & 100 \\
$d/2$   & $\arg\min_t \nabla \beta_0(t)$ & $0.838$ & $0.893$ & 100 \\
$d/2$   & $\arg\max_t \nabla \beta_1(t)$ & $1.172$ & $0.928$ & 100 \\
$d/2$   & $\frac{1}{2}\!\left(\arg\min_t \nabla \beta_0(t)+\arg\max_t \nabla \beta_1(t)\right)$ & $1.005$ & $0.953$ & 100 \\
$h/2$   & $\arg\min_t \nabla \beta_1(t)$ & $1.103$ & $0.859$ & 100 \\
\hline
\end{tabular}
\end{table}

\subsubsection{Open-pore 2D structures}\label{sec:results_lengthscales_open}

In open-pore structures, the dictionary of Sec.~\ref{sec:results_lengthscales} predicts the same readouts for $r$ and $\frac{h}{2}$ as in closed-pore structures, while the predictors which predict $\frac{d}{2}$ in closed-cell structures now serve to predict half the throat width $\frac{w}{2}$. 

\begin{figure}[ht]
 \centering
 \includegraphics[width=0.9\linewidth]{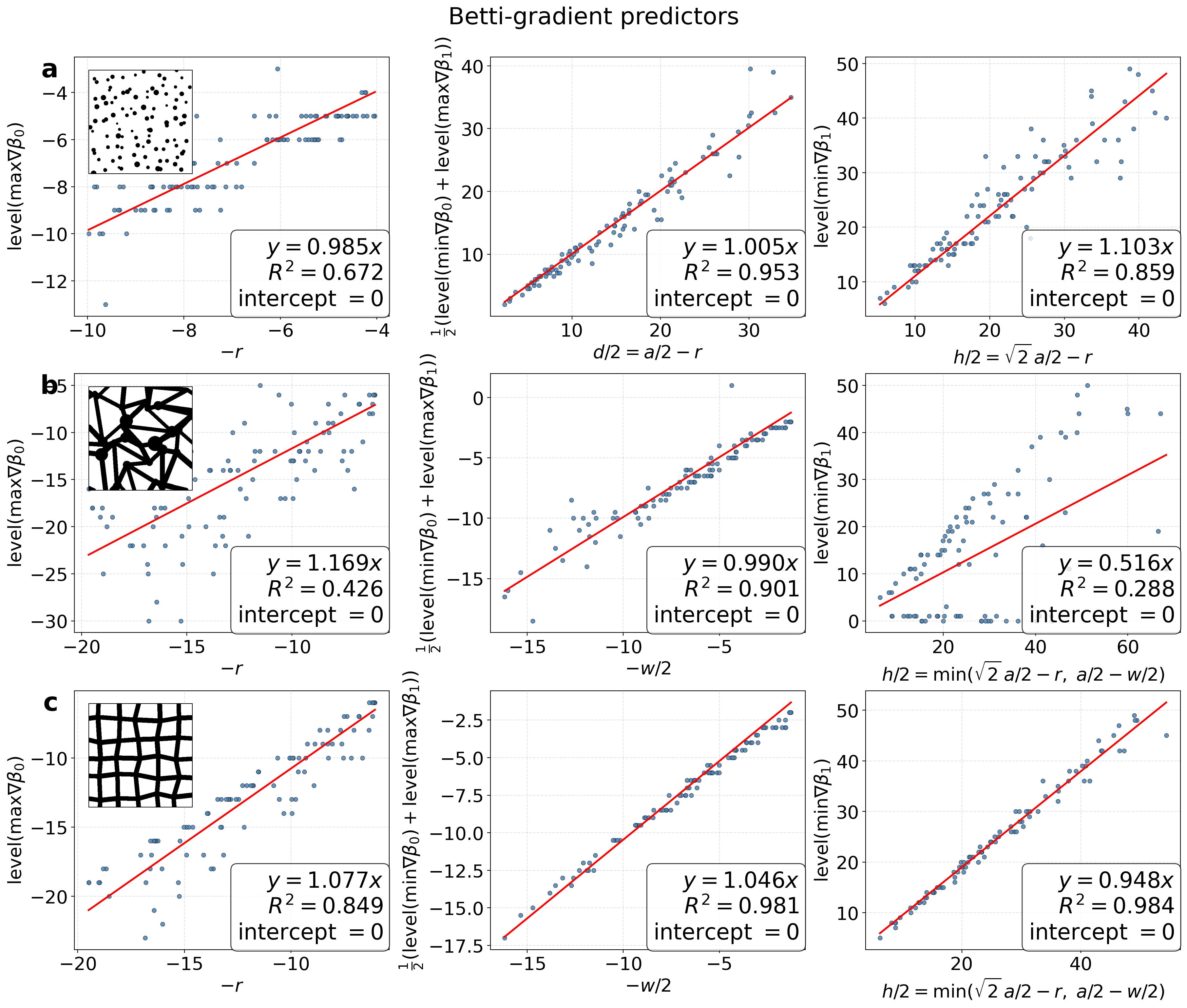}
 \caption{The relationship between Betti-curve based predictors and characteristic length scales in datasets with (a) closed cells and high noise and open cells with high noise (b) and open cells with low noise (c).}
 \label{fig:Betti_vs_shape}
\end{figure}

As for the case of closed-cell structures, we introduce realistic noise and use the first of the 2 datasets of $100$ strongly randomised open-celled structures described in Sec.~\ref{sec:the-open-network-dataset}. Before plotting Betti curves and computing their gradients, we compute persistence intervals for the relevant topological features and remove all intervals with persistence $<1.5$, which we found necessary to remove artefacts of voxelisation. As in the closed-cell dataset, we then fit linear regressions with intercept fixed at $0$ for the predictor-predicted pairs. As seen in the middle row of Fig.~\ref{fig:Betti_vs_shape}, the proposed predictors in practice work much weaker for open-celled than for close-celled structures with a similar level of noise in parameter values. However, in the case of the second of the 2 datasets of $100$ open-celled structures described in Sec.~\ref{sec:the-open-network-dataset}, in which structures have a much lower level of noise in feature positions and sizes, the accuracy of the method is comparable to the case of the dataset with close-celled structures, as presented in the bottom row of Fig.~\ref{fig:Betti_vs_shape}.  

\subsection{Relation to permeability and heat transfer}\label{sec:results_permeability_heat}

\subsubsection{Effective thermal conductivity in 3D porous structures}\label{sec:results_heat_transfer}

We investigate whether there is a systematic relationship between the trace of the heat-transfer tensor and the two openness-related indices. For this purpose, we used the $250$ stochastic three-dimensional pore-solid geometries at $200^3$ voxels described in Sec.~\ref{sec:dataset_for_heat_transfer}, built as five independent ensembles of $50$ samples each, generated with different parameter ranges, mainly to achieve varying average volume fraction across the various subsets (group means of VF varied from roughly $0.31$ to $0.93$, depending on the sampling law). The numerical protocol used to compute the apparent thermal-conductivity tensor is described in Sec.~\ref{sec:methods_simulations}.

Pooling all five ensembles, the Pearson correlation between $\tau$ and $\operatorname{tr}(K)$ was $\approx 0.40$, while the correlation between $\phi_0$ and $\operatorname{tr}(K)$ was $\approx 0.36$, indicating a moderate positive linear association between these openness descriptors and bulk transport capacity under the stated contrast in $k$. The correlation varied substantially between the subsets, with $\tau$ being consistently slightly better predictive of $\operatorname{tr}(K)$ than $\phi_0$ for all studied subsets. Results are summarised in Table~\ref{tab:heat_transfer_correlations}.

\begin{table}[ht]
\centering
\caption{Pearson correlations between openness indices and $\operatorname{trace}(K)$ across the 3D mixed-cells datasets.}
\label{tab:heat_transfer_correlations}
\resizebox{\linewidth}{!}{%
\begin{tabular}{lrrrr}
\hline
\textbf{Subset no.} & \textbf{$n$} & \textbf{Mean filled VF} & \textbf{$\mathrm{corr}\!\left(\tau,\operatorname{trace}(K)\right)$} & \textbf{$\mathrm{corr}\!\left(\phi_0,\operatorname{trace}(K)\right)$} \\
\hline
1  & 50 & 0.500374 & 0.415415 & 0.313777 \\
2  & 50 & 0.314494 & 0.342378 & 0.261867 \\
3  & 50 & 0.857354 & 0.484695 & 0.435153 \\
4  & 50 & 0.929096 & 0.321473 & 0.261911 \\
5  & 50 & 0.944074 & 0.606710 & 0.347216 \\
combined & 250 & 0.763078 & 0.396557 & 0.362976 \\
\hline
\end{tabular}%
}
\end{table}

\subsubsection{Permeability in 3D porous structures}\label{sec:results_permeability}

We continue our investigation of the relationship between $\tau$ and measurable physical variables. We use for this purpose the reported permeabilities and binarized structure shapes in 11 X-ray investigated sandstone structures reported in the dataset of Neumann et al. \cite{Neumann2020}. 

In that case we find no meaningful correlations between $\tau$ itself and permeability. However, we hypothesise that permeability might be strongly influenced not only by the number of loops formed by channels in the porous rock, but also by their maximal and minimal widths. Thicker loops, which would therefore manifest at lower filtration levels, should contribute to increased flow permeability more. Therefore, we suggest the sums over the filtration levels (a) $\sum_{-\infty}^{-1} \tau(t) dt$ and (b) $-\sum_{-\infty}^{-1} t \tau(t) dt$ as possible variables to correlate with permeability. Here $\tau(t)$ is a generalisation of $\tau$ in that the loops and connected components are not evaluated on the original image, but on a dilated/eroded image, on any filtration level $t$. The first of them would weight more strongly porous loops of large width since they would be counted multiple times in the course of the integration, while the latter would further increase that effect by weighing the loops by thickness explicitly.   

\begin{figure}[ht]
 \centering
 \includegraphics[width=0.8\linewidth]{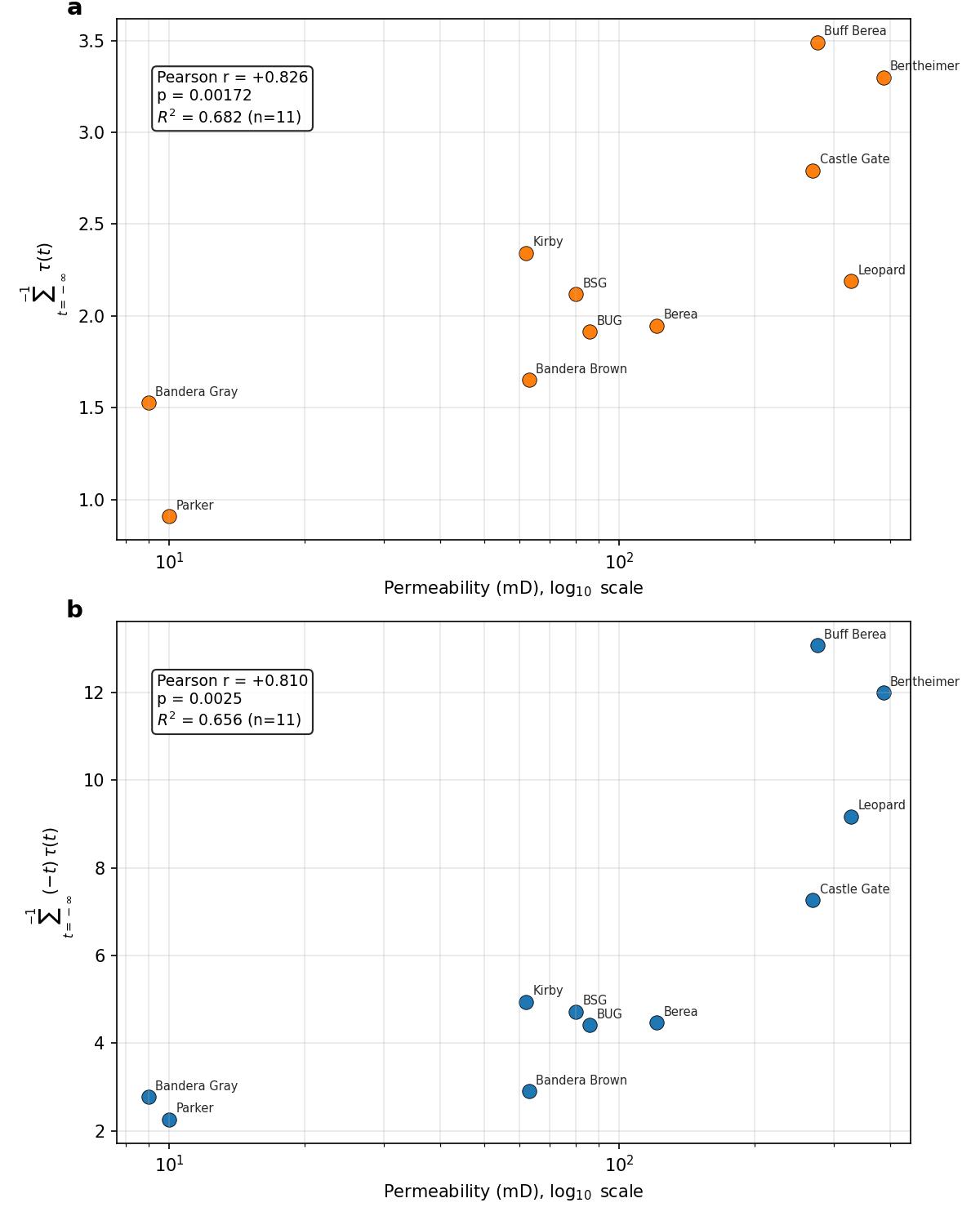}
 \caption{Relationships between sums based on $\tau$ and $\log_{10}$(permeability) in a dataset of sandstones\cite{Neumann2020}. The datapoint descriptions correspond to original sample names.}
 \label{fig:betti_observables_vs_permeability}
\end{figure}

We implement a computation of these sums and find that both have strong and similar Pearson correlations with log-scale permeability (see Fig. \ref{fig:betti_observables_vs_permeability}). Subject to further investigations, we suggest the integrals could potentially serve to screen for candidate structures with strong permeability.    

\section{Discussion}\label{sec:discussion}

Across the synthetic 2D and 3D systems considered here, the cell-openness index $\tau$ behaves as a topological complement to the numerical open porosity $\phi_0$. The two indices are strongly correlated through an approximate power law $\phi_0 \propto \tau^{1/k}$, with $k\approx 2$-$3$ in 2D and $k\approx 3$-$5$ in 3D, but $\tau$ retains discriminatory power in regimes where $\phi_0$ saturates close to unity. The mismatch between the two indices in that regime carries genuine structural meaning: it flags pore networks in which most components individually reach the sample boundary but are mutually disconnected from one another- a feature that pycnometry, by construction, cannot resolve.

Betti curves computed on signed-distance filtrations carry predictive length-scale information on the 2D datasets with both open and closed pores. For these datasets, the points with extremum gradients of $\beta_0$ and $\beta_1$ are the basis of effective predictors of the mean pore radius $r$, the halfwidth of the inter-pore distance $d/2$, the half-thickness of the solid cores $h/2$ and the half-width of throats between pore centers $w/2$, with slopes close to $1$ and $R^2$ between $0.67$ and $0.95$. However, for the open-celled structures the predictors lose their accuracy at lower levels of noise than for close-celled structures. This observation somewhat resembles the conclusions of Robins et. al \cite{Robins2016} about predicting percolation radii based on topological persistence data; their study concluded that it was much more difficult for highly heterogenous systems.


More research is needed into both the type of structural information that $\tau$ can provide and what physical properties it can help predict. However, the work so far suggests that $\tau$ is at least as informative as $\phi_0$- the fraction of open pores measured by gas pycnometry- the reporting of which is an industry standard for commercially available porous materials. We suggest that where plausible (specifically, where high-quality 3D imaging of a porous structure is available), our cell-openness index should be reported alongside the fraction of open pores measured by gas pycnometry.

\section*{Acknowledgements}
\addcontentsline{toc}{section}{Acknowledgements}

“This research was funded in whole or in part by the National Science Centre, Poland, grant number 2021/03/Y/ST5/00232. For the purpose of Open Access, the author has applied a CC-BY public copyright licence to any Author Accepted Manuscript (AAM) version arising from this submission.

Financial support from the PORMETALOMICS project, funded by the
National Science Centre, Poland (project no. 2021/03/Y/ST5/00232)
within the M-ERA.NET 3 programme, is gratefully acknowledged.
This project has received funding from the European Union’s Horizon
2020 research and innovation programme under grant agreement No
931174.”

The authors used Anthropic's Claude.AI and OpenAI's ChatGPT to assist with language editing,
spellchecking, and paraphrasing of selected parts of the manuscript and Cursor.AI for assistance with figure generation.
These tools were not used to generate scientific ideas, data, analyses,
results, or references. All scientific content,
interpretations, and conclusions were developed and verified by the
authors, who take full responsibility for the final manuscript.

The authors thank Dr. Maciej Matyka for insightful discussions.

The manuscript was published under Open Access in Transport Phenomena and is available under the link https://www.degruyterbrill.com/document/doi/10.1515/tp-2026-0074/html.

\bibliographystyle{plain}
\nocite{*}
\bibliography{references}

\end{document}